# Nanotube ferroelectric tunnel junctions with giant tunneling electroresistance ratio

Jiu-Long Wang, Yi-Feng Zhao, Wen Xu, Jun-Ding Zheng, Ya-Ping Shao, Wen-Yi Tong[*], Chun-Gang Duan[*]

**ABSTRACT :** Low-dimensional ferroelectric tunnel junctions are appealing for the realization of nanoscale nonvolatile memory devices due to their inherent advantage of device miniaturization. Those based on current mechanisms still have restrictions including low tunneling electroresistance (TER) effects and complex heterostructures. Here, we introduce an entirely new TER mechanism to construct the nanotube ferroelectric tunnel junction with ferroelectric nanotubes as the tunneling region. When rolling a ferroelectric monolayer into a nanotube, due to the coexistence of its intrinsic ferroelectric polarization with the flexoelectric polarization induced by bending, there occurs metal-insulator transition depending on radiative polarization states. For the pristine monolayer, its out-of-plane polarization is tunable by an in-plane electric field, the conducting states of the ferroelectric nanotube can thus be tuned between metallic and insulating via axial electric means. Using α-$In_2Se_3$ as an example, our first-principles density functional theory calculations and nonequilibrium Green's function formalism confirm the feasibility of the TER mechanism and indicate an ultrahigh TER ratio exceeding $9.9 \times 10^{10}\%$ of the proposed nanotube ferroelectric tunnel junctions. Our findings provide a promising approach based on simple homogeneous structures for high density ferroelectric microelectric devices with excellent ON/OFF performance.

**Keywords:** *ferroelectric nanotube, flexoelectric effect, metal-insulator transition, ferroelectric tunnel junction*

# Introduction

Due to its potential in nonvolatile functional devices, ferroelectricity is a hot issue over the last decades.[1, 2] Low-dimensional van der Waals (vdW) ferroelectric materials have benefits over their three-dimensional counterparts in terms of lower energy consumption and higher response speed,[3-11] providing a more promising platform for ferroelectric device applications. In this regard, atomic thickness ferroelectric tunnel junction (FTJ),[12-17] with the inherent advantage of device miniaturization, have been proposed.

Two-dimensional (2D) FTJ constructed by 2D ferroelectric materials with in-plane polarization, including Group-IV monochalcogenides[10, 11, 18] and elemental Group V buckled monolayers,[19] are pioneers. Shen et al. designed a SnSe-based homogeneous FTJ with p- and n-type doping at the two terminals, where both barrier width and height are dynamically manipulated during ferroelectric switching.[14] Similar tunneling electroresistance (TER) effects have been realized in in-plane ferroelectric monolayers sandwiched between leads doped with holes and electrons respectively.[17, 20] These studies are based on asymmetric screening effects in the two electrodes, belonging to the general TER mechanism.[21] Another strategy is implemented through the intrinsic out-of-plane (OOP) polarization of 2D ferroelectrics, with $In_2Se_3$ and $CuInP_2S_6$ as representatives.[22-27] In a vertical heterostructure consisting of a ferroelectric monolayer and another vdW one on top or bottom, a transverse electric field tunes the out-of-plane polarization with the change of charge transfer between two monolayers, making metal-insulator transition feasible. As demonstrated in $α$-$In_2Se_3$/SnTe[13] and graphene/$α$-$In_2Se_3$[15] systems, such in-plane 2D FTJ characterized by the control of conducting states between metallic and insulating nature generally produce an extremely high TER ratio. We intend to propose a new TER mechanism by introducing a ferroelectric nanotube. As the representative one-dimensional structures, nanotubes with typical bending structures exhibit flexoelectric effect: an unconventional electromechanical coupling that occurs under inhomogeneous mechanical or electrical conditions.[28-32] In addition to the well-studied carbon nanotubes and those built on transition metal dichalcogenides,[33-36] all 2D materials, in principle, could be made into corresponding nanotubes.[37-39] Ferroelectric nanotubes, natively yielding the coexistence of flexoelectric effect and ferroelectricity, are however rarely reported.[36, 40] When a ferroelectric monolayer with OOP polarization is rolled into a nanotube, the original OOP ferroelectricity becomes radiated and might be tunable by switching the axial electric field. Depending on whether the flexoelectric and radiated ferroelectric directions are the same or opposite, two inequivalent conducting states are expected, which offer the chance to construct nanotube FTJ.

Following the strategy, we take the α-In$_2$Se$_3$ with intrinsic OOP polarization at room temperature as an example.[22-24] Due to its asymmetric structure, the freestanding 2D sheets should be unstable. Han et al. has already experimentally established opportunities for nanoscale bending with high angles in α-In$_2$Se$_3$.[41] Our first-principles calculations confirm that In$_2$Se$_3$ monolayer tends to naturally wrap and form ferroelectric nanotube with a well-defined size. When the radiative polarization switches from the inward to the outward direction controlled by an axial electric field, due to its joint effect with the flexoelectric polarization, conduction band shift to lower energy level while valance band shift to higher one, inducing insulator-metal transition in In$_2$Se$_3$ nanotubes. Transport calculations further confirm the existence of obvious current difference between In$_2$Se$_3$ nanotubes in two polarization states, suggesting the feasibility of nanotube FTJ with an ultra-high TER ratio of $9.9 \times 10^{10}\%$. The flexoelectric-related TER mechanism here is clearly in contrast to previous studies of 2D FTJs.[13-15] Its homogeneous nature is advantageous for potential manufacture as well.

## Computational methods

The density functional theory (DFT) calculations are carried out by using the Vienna Ab initio Simulation Package (VASP)[42] with a plane wave basis set and the projector-augmented wave (PAW)[43] method to geometry optimizations and electronic structure calculations of slab models. The exchange correlation functional is treated in generalized gradient approximation (GGA) with the type of Perdew−Burke−Ernzerhof (PBE).[44] The kinetic-energy cutoff of 500 eV is applied to the plane wave expansion and a Γ-centered 8 × 1 × 1 k points grid is adopted for Brillouin zone sampling. All the structures are optimized until the Hellmann−Feynman forces are below 1 meV/Å, and the convergence threshold of electronic energy is $10^{-6}$ eV. At least 15 Å vacuum space is applied to avoid interactions except for the direction of

transportation. The vdW correction incorporated using the DFT-D3 method.[45]

The device properties of the $In_2Se_3$ nanotube are calculated by using DFT plus nonequilibrium Green's function formalism (DFT+NEGF approach),[46, 47] as carried out in the OPENMX software package.[48] A real-space mesh cutoff energy of 200 Ryd is used to guarantee the good convergence of the device configuration. The residual minimization method in the direct inversion iterative subspace with Kerker metric (RMM-DIISK) is applied to help charge mixing.

The electron temperature is set at 300 K, and $1 \times 10 \times 10$ k mesh is used for the self-consistent calculations to eliminate the mismatch of Fermi level between electrodes and the central region. The spin-resolved conductance is obtained by the Landauer−Buttiker formula[49] at equilibrium state:

$$G_0 = \frac{e^2}{h} \sum_{k_{||}} T_\sigma(k_{||}, E_F), \quad (1)$$

and the current is defined as:

$$I = \frac{e^2}{h} \int T_\sigma(E)[f(E - \mu_l) - f(E - \mu_r)] dE \quad (2)$$

where $T_\sigma(k_{||}, E_F)$ is the transmission coefficient with spin $\sigma$ at the transverse Bloch wave vector $k_{||}$ and Fermi energy $E_F$, $e$ is the electron charge and $h$ is the Planck constant, $f$ is the Fermi distribution function and $\mu_l$ and $\mu_r$ are the chemical potentials of the left and right leads, respectively. At nonequilibrium state, the current is calculated as the integral.

**Results and discussion**

As illustrated in Fig.1, we focus on the ferroelectric α-$In_2Se_3$ monolayer with intrinsic OOP polarization, whose polarization states are determined by the relative location of tetrahedral and octahedral coordination, and can be controllable by an in-plane electric field. Following the experiment that successfully bending it along the zigzag direction,[41] we construct a zigzag nanotube (ZN) with 30 unit cells of α-$In_2Se_3$ monolayer (the blue box in Fig.2(a)), which should be large enough to avoid its structural collapse. As shown in Fig.1, the 2D OOP polarization is transformed to 1D radiative polarization with two states: P-out and P-in, when we bending α-$In_2Se_3$ monolayer.

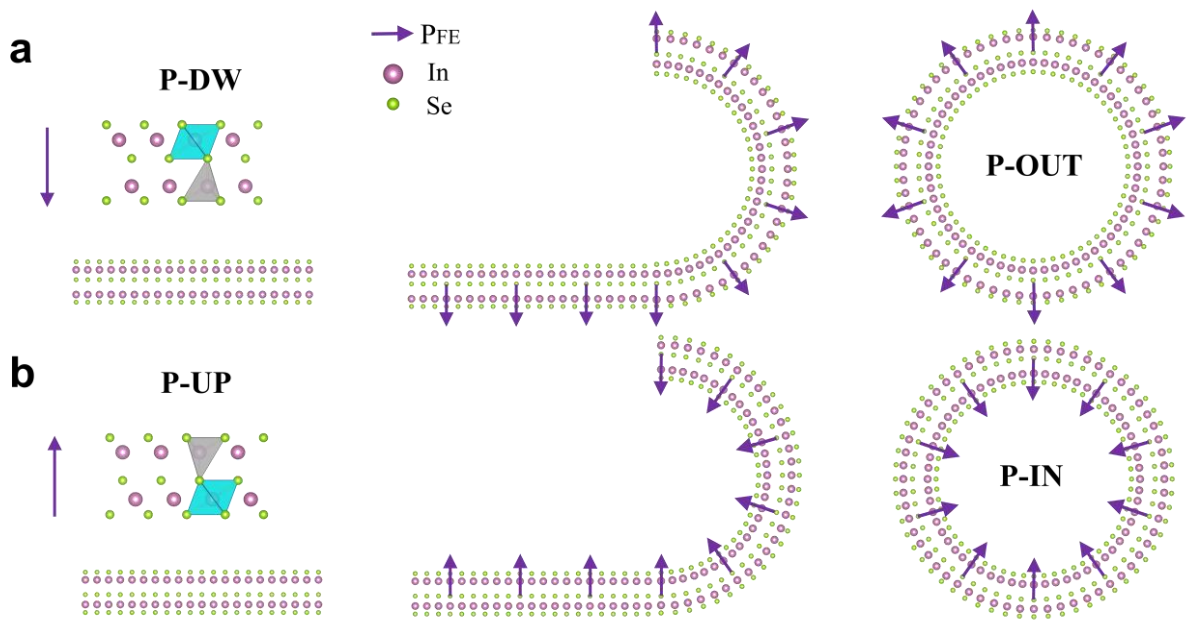

**Figure 1**. (a,b) The tuning progress of OOP ferroelectric polarization to radiative polarization. Violet (green) balls indicate In (Se) atoms, and the violet arrows represent the intrinsic ferroelectric polarization. Gray (wathet) polyhedral indicate tetrahedral (octahedral) coordination.

When rolls up, the outside Se2-In2 bonds are extended, while the inside Se1-In1 bonds are compressed. This strain gradient inwardly propels middle-sited Se3 to simultaneously elongate(shorten) its bonding with In-2(In-1), stabilizing both the outside tetrahedral and the inside octahedral coordination. Such kind of movement trend identifies the flexoelectric polarization pointing outward. The $In_2Se_3$ nanotube, then, could possess two different polarization states, that is the polarization-out (P-out) and the polarization-in (P-in) one. As shown in Fig.2(b), the flexoelectric polarization and intrinsic ferroelectric polarization of the P-out state are in the same radial direction. Such a combination effect causes its larger effective polarization, reflecting in a bigger displacement between the center of In cations and Se anions $\delta_{In-Se}$ with the magnitude of 0.30 Å. The P-in state, reversely, corresponds to the compensation of the two polarizations with opposite directions, where the net polarization is relatively small ($\delta_{In-Se}$ = -0.14 Å). When the polarization switches from outward to inward, negatively charged $Se^{2-}$ ions move to the opposite direction of the additional inward field, resulting in the larger inner $D_{in}$ (from 32.40 to 32.83Å) and outer diameters $D_{out}$ (from 46.29 to 46.47Å) of the P-in state. Compared with the magnitude of 13.89 Å in the P-out state, the difference between the outer and inner diameters of the P-in state is smaller ~13.64 Å, in consistent with its little $\delta_{\text{In-Se}}$.

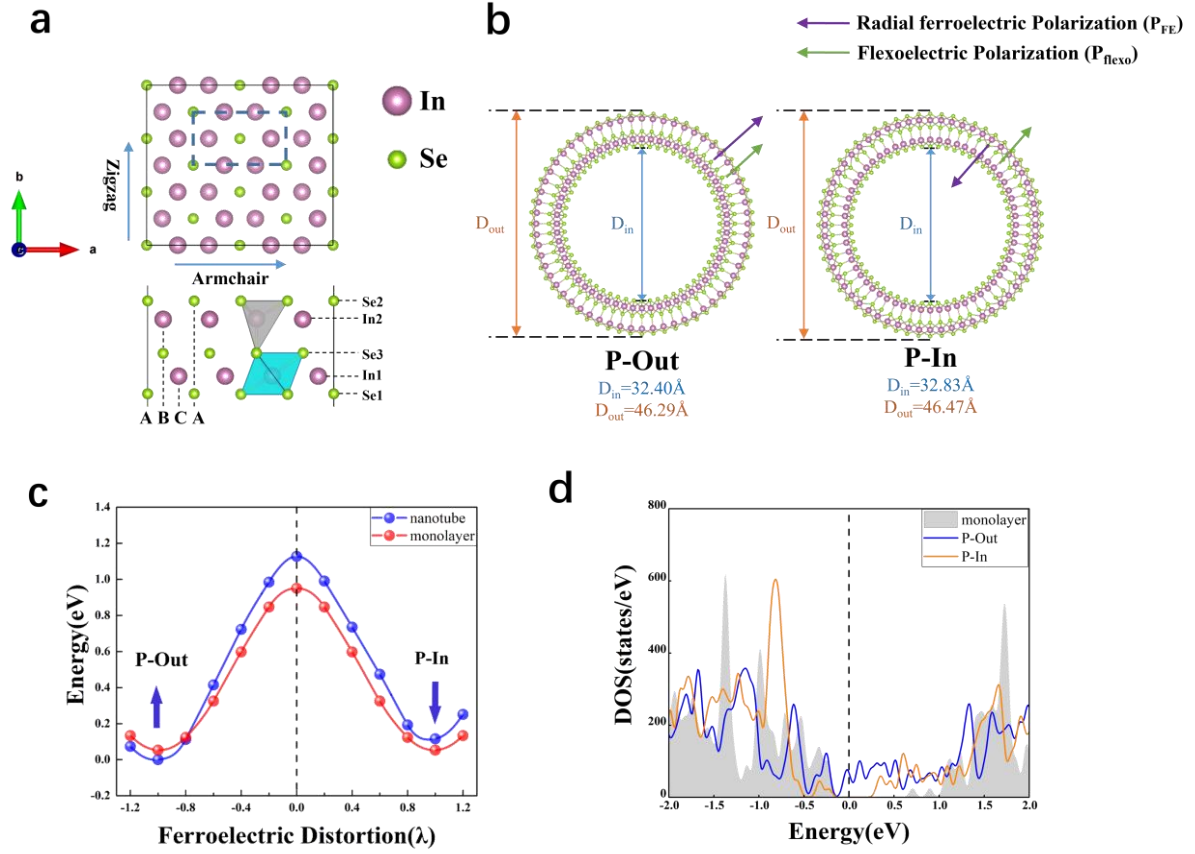

**Figure 2**. (a) Schematic diagram of α-$In_2Se_3$ monolayer. The blue box represents the unit cell. (b) The crystal structure of P-out and P-in zigzag nanotubes. The green arrows represent the orientation of the radiative flexoelectric polarization, while the violet arrows correspond to that of the intrinsic ferroelectric polarization. We define inner diameter ($D_{in}$) as the diameter of the inner Se layer, and outer diameter ($D_{out}$) as the diameter of the outer Se layer. (c) Potential energy profile as a function of ferroelectric distortions in an $In_2Se_3$ monolayer with the red line and its corresponding ZN with the blue line. The ferroelectric displacements λ are normalized so that λ = +1 and −1 correspond to the P-in/P-down and P-out/P-up states, respectively. (d) The density of state (DOS) of monolayer $In_2Se_3$ P-out and P-in ZN.

To confirm the stability of such ferroelectric $In_2Se_3$ ZN, we calculate the strain energy of both P-out and P-in states. The strain energy is defined as the difference between the energy of the nanotube and the corresponding 2D sheet with the formula:[50]

$$E_{\text{strain}} = \frac{E_{\text{tube}}}{N_{\text{tube}}} - \frac{E_{\text{sheet}}}{N_{\text{sheet}}}, \qquad (3)$$

where $E_{tube}$ and $E_{sheet}$ are the energy of nanotube and the energy of the corresponding 2D ferroelectric In$_2$Se$_3$ sheet, respectively. $N_{tube}$ and $N_{sheet}$ mean the number of atoms. As expected, we gain a negative $E_{strain}$ of -5.3meV in P-out state, proving the trends of wrapping in α-In$_2$Se$_3$ monolayer. For the P-in state, its positive $E_{strain}$ (17.8meV) indicates there exists preferred wrapping direction to guarantee the same-oriented flexoelectric and ferroelectric polarization. Nevertheless, according to the experimental observation that α-In$_2$Se$_3$ bending maintains its original polarization direction,[41] energetically unfavored P-in state is achievable. Similar as the α-In$_2$Se$_3$ monolayer where transvers electric field works to reverse the out-of-plane polarization via laterally shifting the central Se3 layer,[24] with assistance of a moderate axial electric field, it's possible to switch between P-in and the more stable P-out state through readily accessible kinetic pathways.

Similar as traditional ferroelectric materials like bulk BaTiO$_3$ and PbTiO$_3$,[51] the potential energy profile of the ferroelectric soft mode motion in an In$_2$Se$_3$ sheet is symmetric (the red line in Fig.2(c)). When rolled up into nanotubes, the existence of flexoelectric polarization in a fixed direction as marked in Fig.2(b), however, destroys this symmetric energy profile. Its same direction as the intrinsic polarization causes a combination effect in the P-out state, which yet becomes the partially compensation effect in the P-in state due to the opposite flexoelectric and ferroelectric direction. Indeed, our calculations, as clearly shown from the blue line in Fig.2(c), find two inequivalent energy minima, in consistent with previous results of strain energy. This confirms the stability of two polarization states and is exactly the signature of asymmetric ferroelectricity.

It is interesting to point out that as shown in Fig.2(d), when the polarization is pointing outward, the P-out system is in a metallic phase. When it switches to be inward, the P-in ZN turns to be insulating with a band gap of 0.22eV. As a reference and compared in Fig. 2(d), the monolayer ferroelectric α-In$_2$Se$_3$ is an insulator with an indirect band gap of 0.76eV.[24]

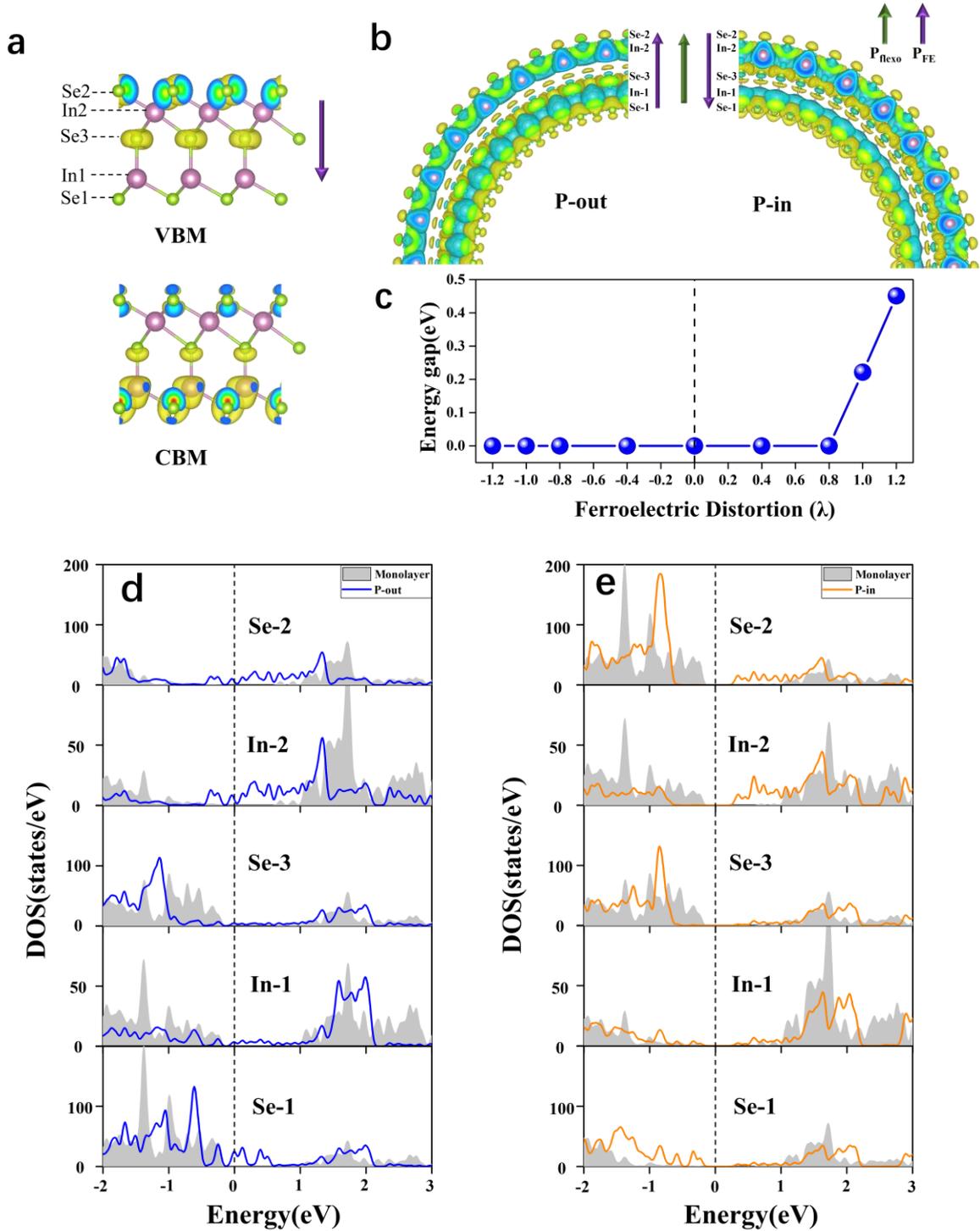

**Figure 3**. (a) Charge denstity of valence band maximun (VBM) and conduction band minimun (CBM) of α-In$_2$Se$_3$ monolayer in real space. (b) The deformation charge density of P-out and P-in ZN, from bottom to top are labeled as Se-1, In-1, Se-3, In-2, Se-2. Yellow and blue regions denote electron accumulation and depletion, respectively. The isosurface value is $1.0 \times 10^{-4}$ e/bohr.[52] Violet arrow represent the orientations of the intrinsic ferroelectric polarization. (c) Energy gap with ferroelectric distortion of In$_2$Se$_3$ nanotube. The ferroelectric displacements λ are normalized so that λ = +1 and −1

correspond to the P-in and P-out states, respectively. (d,e) Layer-dependent DOS of P-out and P-in ZN. As reference, those of the monolayer are displayed with gray zones.

To deeply understand the mechanism of the metal-insulator transition, we firstly investigate the layer-dependent density of states (DOS) of monolayer. Due to the presence of a built-in electric field $E_{FE}$ generated by intrinsic ferroelectric polarization $P_{FE}$, the electrons are transferred from the negatively poled surface to the positive one, forming the electrostatic potential difference $\Delta\Phi_1$ between two opposite surfaces. As shown in Fig.3(a) and diagrammed in Fig.4(a), the conduction band minimum (CBM) is then mainly occupied by the states from the positively poled surface while the valence band maximum (VBM) corresponds to those from the opposite region. Note that the middle Se, in the tetrahedral and octahedral coordination with both sides, contributes to them as well. An additional electric field $E_{flexo}$ generated by the flexoelectric effect could enhance or weaken the potential difference depending on the $P_{FE}$ direction.

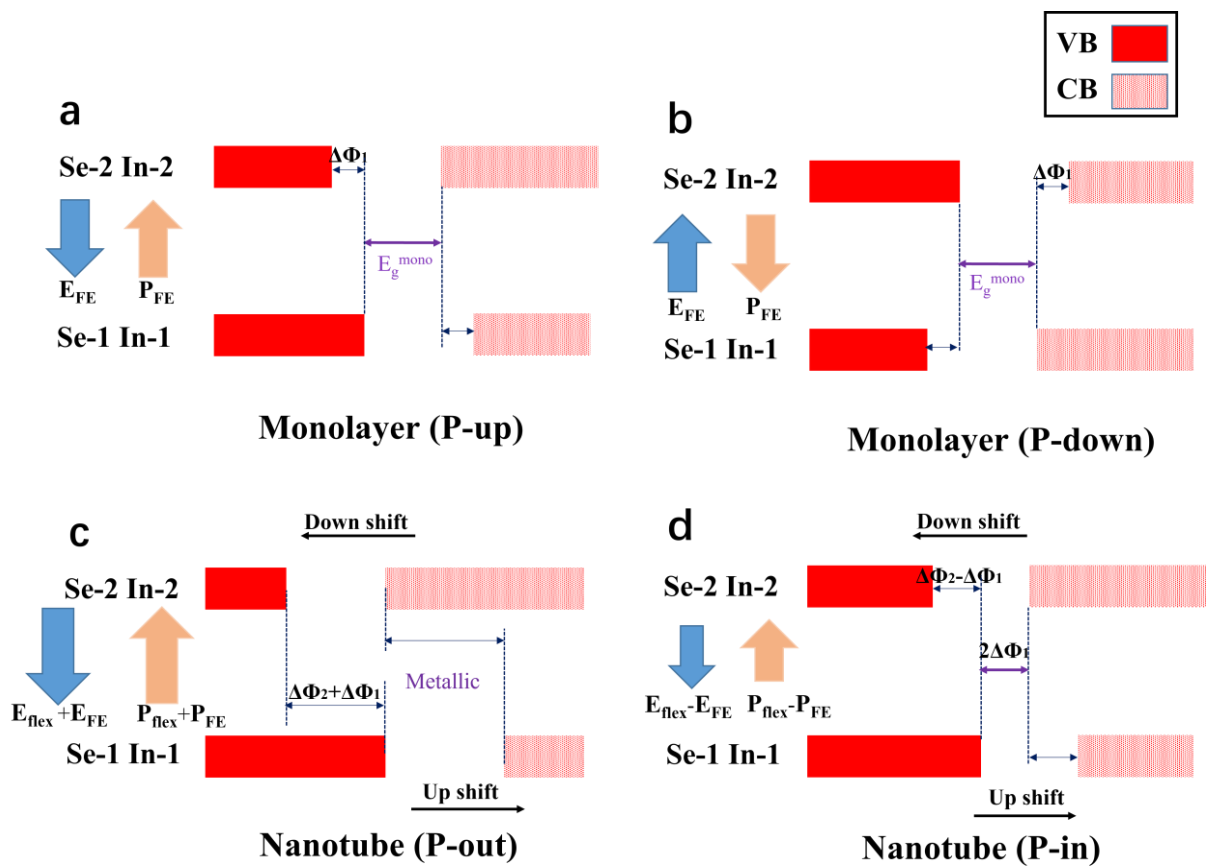

**Figure 4**. Band alignment diagrams of In$_2$Se$_3$ monolayer with polarization pointing (a) upward and (b) downward, and nanotubes in the (c) P-out and (d) P-in states.

The rolling of the monolayer induces an effective outward flexoelectric polarization as analyzed before. When In$_2$Se$_3$ polarization points outward as the flexoelectric polarization direction, the electronic levels from the outside Se-2, In-2 go down, reinforcing the downshift of the CBM. Meanwhile, an additional upshift of the VBM occur due to the further upward moving of the inside Se-1, In-1 states. As the enhanced electrostatic potential difference $\Delta\Phi_2$ is larger than the original band gap of monolayer $E_g^{mono}$, the gap is eventually closed and renders the metallic P-out state (see Fig.4(d) and Fig.4(c)). However, when ferroelectric polarization points inward, $E_{flexo}$ counteracts the opposite $E_{FE}$, pulling the originally higher outside Se-2, In-2 states downward and pushing the lower Se-1, In-1 states upward (Fig.3(e)). Within the same $\Delta\Phi_2$ as the P-out state, the P-in state is expected to maintain its insulating character with the band gap of $2\Delta\Phi_1$ (see Fig.4(d)). The additional electrostatic potential difference needed to close the gap here should be greater than $E_g^{mono}+2\Delta\Phi_1$, indicating its hysteretic nature of metallicity compared with the P-out state. Regarding the relatively free Se-2 layer, they could react the additional flexoelectric field through the ionic displacement to stabilize the system, leading to uniformly downward movements of their energy levels (Fig.3(d-e) middle panels). It is also worth noting that for the P-out states (left side of Fig.3(c)), the metallicity should be robust. Within smaller ferroelectric polarization, $E_g^{mono}$ decreases, $\Delta\Phi_2$ generated by the fixed flexoelectric field is easier to compensate $E_g^{mono}$, making the nanotubes metallic. For the P-in states, similarly, the smaller the ferroelectric polarization is, VBM and CBM are more likely to translate toward the Fermi level, which is confirmed by the right side of Fig.3(c).

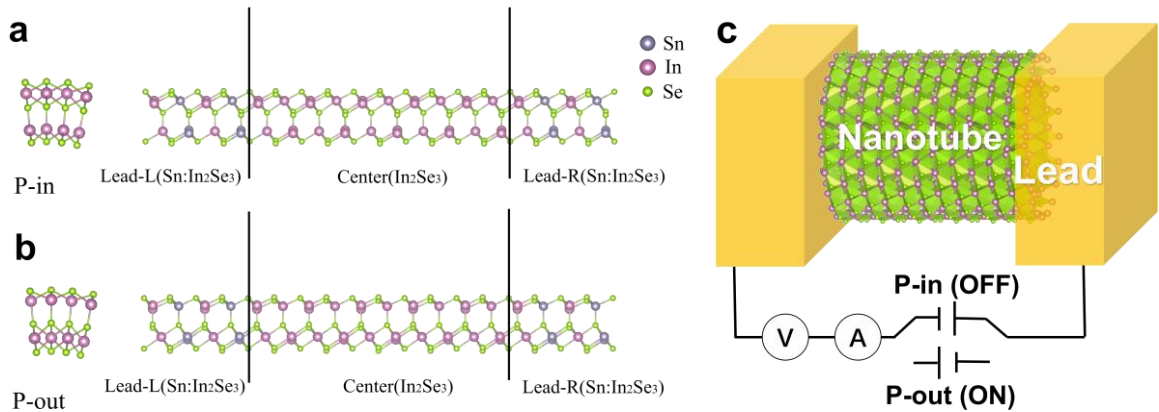

**Figure 5**. (a,b) Side view of P-in and P-out nanotube unit and axial transportation structures. Sn doping In$_2$Se$_3$ are taken as leads, where gray balls represent Sn atoms. (c) Schematic diagram of nanotube FTJ. Au leads are attached to both sides of the nanotube to create a positive or negative axial electric field, which can reverse the radiative polarization of the nanotube in the middle.

To explore the feasibility of In$_2$Se$_3$ nanotube in microelectronic application, we further study the transport properties of ZN. Units of P-in and P-out nanotubes were utilized, with Sn doping α-In2Se3 acting as leads. The conductance of the P-in and P-out nanotube unit cell under 0 bias were measured to be $9.14\times10^{-11}$ *S* and $9.10\times10^{-2}$ *S*, respectively. Additionally, a transport model was established along the radial direction, as shown in the supplementary information (Fig.S4), and similar results were obtained. Therefore, regardless of whether the bias is applied along the axial or radial direction, the transport characteristics remain almost the same, demonstrating the potential of α-In$_2$Se$_3$ nanotubes in FTJ devices.

Taking into account the dramatic difference in the transport characteristics between P-in and P-out states of In$_2$Se$_3$ ZN and its tunable radiative polarization under an axial electric field, we design a nanotube FTJ, as diagrammed in Fig.5(c). Through controlling the biswitch to generate axially positive or negative bias along the nanotube channel region, we could electrically reverse the radiative polarization of the α-In$_2$Se$_3$ ferroelectric nanotube, realizing the non-volatile storage of the binary information. The ZN in the P-in state is still a semiconductor as the pristine sheet. Therefore, almost no electrons go through the FTJ, leading to the "OFF" state with low transmission coefficient. In contrast, the metallic P-out state corresponds to the "ON" state within a giant current flow. To assess the TER effect in the nanotube FTJ, we define the TER ratio as:

$$\text{TER} = \frac{G_{out}-G_{in}}{G_{in}}, \tag{4}$$

where $G_{out}$ and $G_{in}$ represent the conductance of P-out and P-in ZN respectively. our FTJ exhibits an ultrahigh TER ratio of $9.9\times10^{10}\%$, which is comparable to previous reports based on the control of conducting states[13, 15] and greatly exceeds the 2D FTJs belonging to the general TER mechanism.[14, 16, 17]

## Conclusion

In summary, we propose a nanotube FTJ with rolling 2D ferroelectric sheets as the central tunneling region, whose radiative polarization is reversible via axial electric means. The TER mechanism here roots in the tunability of ferroelectric nanotubes from metal to insulator, which is closely related to the hysteretic behavior of metallicity when the additional electric field generated by the flexoelectric effect is opposite to the one from the intrinsic ferroelectric polarization. We adopt $α$-$In_2Se_3$ as an example to construct the ferroelectric nanotube. Its switching between inward and outward radiative polarization under bending has been proved in experiment very recently.[41] Based on such a ferroelectric nanotube, we predict the existence of a giant TER effect in the nanotube FTJ with the TER ratio of up to $9.9 \times 10^{10}$%, which is several orders of magnitude greater than traditional FTJs with 2D in-plane ferroelectrics as tunnel barriers. When compared to other FTJs with the change of conducting states between metallic and insulating nature, our mechanism benefits a comparable TER value, while its homogeneous structure, rather than vertical heterostructures, makes it advantageous in practical applications. We note that the metal-insulator transition does not exist in $α$-$In_2Se_3$ alone. It is, in principle, versatile in ferroelectric nanotubes with out-of-plane polarization. Moreover, within the coexistence of the ferroelectric polarization and flexoelectric one, similar phenomena could be reproduced in a variety of systems and geometry structures. The application of this effect is not restricted in nanotube FTJ with an extremely high TER ratio as well. Such kind of versatile strategy to realize metal-insulator transition may provide a compelling toolbox for developing advantageous ferroelectric memory and microelectronic devices.

**AUTHOR INFORMATION**
**Corresponding Author**
**Wen-Yi Tong** - Key Laboratory of Polar Materials and Devices (MOE), Ministry of Education, Department of Electronics, East China Normal University, Shanghai, 200241, China
Shanghai Center of Brain-inspired Intelligent Materials and Devices, East China Normal University, Shanghai 200241, China
E-mail: wytong@ee.ecnu.edu.cn
**Chun-Gang Duan** - Key Laboratory of Polar Materials and Devices (MOE), Ministry of Education, Department of Electronics, East China Normal University, Shanghai, 200241, China
Shanghai Center of Brain-inspired Intelligent Materials and Devices, East China Normal University,

Shanghai 200241, China

E-mail: cgduan@clpm.ecnu.edu.cn

**Authors**

**Jiu-Long Wang -** Key Laboratory of Polar Materials and Devices (MOE), Ministry of Education, Department of Electronics, East China Normal University, Shanghai, 200241, China

Shanghai Center of Brain-inspired Intelligent Materials and Devices, East China Normal University, Shanghai 200241, China

**Yi-Feng Zhao -** Key Laboratory of Polar Materials and Devices (MOE), Ministry of Education, Department of Electronics, East China Normal University, Shanghai, 200241, China

Shanghai Center of Brain-inspired Intelligent Materials and Devices, East China Normal University, Shanghai 200241, China

**Wen Xu -** Key Laboratory of Polar Materials and Devices (MOE), Ministry of Education, Department of Electronics, East China Normal University, Shanghai, 200241, China

Shanghai Center of Brain-inspired Intelligent Materials and Devices, East China Normal University, Shanghai 200241, China

**Jun-Ding Zheng -** Key Laboratory of Polar Materials and Devices (MOE), Ministry of Education, Department of Electronics, East China Normal University, Shanghai, 200241, China

Shanghai Center of Brain-inspired Intelligent Materials and Devices, East China Normal University, Shanghai 200241, China

**Ya-Ping Shao -** Key Laboratory of Polar Materials and Devices (MOE), Ministry of Education, Department of Electronics, East China Normal University, Shanghai, 200241, China

Shanghai Center of Brain-inspired Intelligent Materials and Devices, East China Normal University, Shanghai 200241, China

**Author Contributions**

W.Y.T. and C.G.D conceived the idea and supervised the work. J.L.W carried out non-equilibrium Green's function formalism and first-principles calculations. W.Y.T and J.L.W did the data analysis. All the authors discussed the results and co-wrote the manuscript.

**Acknowledgements**

This work was supported by the National Key Research and Development Program of China (Nos. 2022YFA1402902 and 2021YFA1200700), the NSF of China (No. 12134003), Shanghai Science and Technology Innovation Action Plan (No. 21JC1402000), ECNU Multifunctional Platform for Innovation.